# Clustering versus Statistical Analysis for SCA: when Machine Learning is Better

Marcin Aftowicz[1], Ievgen Kabin[1], Zoya Dyka[1], Peter Langendoerfer[1,2]
[1]IHP – Leibniz-Institut für innovative Mikroelektronik, Frankfurt (Oder), Germany
[2]BTU Cottbus-Senftenberg, Cottbus, Germany
{aftowicz, kabin, dyka, langendoerfer}@ihp-microelectronics.com

**Abstract**—Evaluation of the resistance of implemented cryptographic algorithms against SCA attacks, as well as detecting of SCA leakage sources at an early stage of the design process, is important for an efficient re-design of the implementation. Thus, effective SCA methods that do not depend on the key processed in the cryptographic operations are beneficially and can be a part of the efficient design methodology for implementing cryptographic approaches. In this work we compare two different methods that are used to analyse power traces of elliptic curve point multiplications. The first method *the comparison to the mean* is a simple method based on statistical analysis. The second one is *K-means* – the mostly used unsupervised machine learning algorithm for data clustering. The results of our early work showed that the machine learning algorithm was not superior to the simple approach. In this work we concentrate on the comparison of the attack results using both analysis methods with the goal to understand their benefits and drawbacks. Our results show that *the comparison to the mean* works properly only if the scalar processed during the attacked $kP$ execution is balanced, i.e. if the number of '1' in the scalar $k$ is about as high as the number of '0'. In contrast to this, *K-means* is effective also if the scalar is highly unbalanced. It is still effective even if the scalar $k$ contains only a very small number of '0' bits.

## I. Introduction

Elliptic Curve Cryptography (ECC) is a kind of public key cryptography. Each participant has a pair of cryptographic keys – a public key that is a point of an EC over a finite field, and a private key that is a long binary number. There exist standardized protocols for sharing a secret (key exchange approaches), authentication of participants as well as digital signatures based on ECC. Due to the relatively short length of cryptographic keys, the Elliptic Curve (EC) approaches can successfully be applied for implementing confidentiality of communication, data integrity and other security features on devices with constrained resources. ECC operations can be accelerated additionally if implemented in hardware as specialized ASICs, for example for authentication or digital signature generation and/or verification approaches. Such ASICs can reduce the energy consumption of cryptographic operations significantly. This makes ECC attractive for WSN and IoT applications.

The main operation in ECC is the EC point multiplication, denoted as $kP$. A $kP$ operation is a multiplication of an Elliptic Curve point $P$ by a scalar $k$. Point $P$ is expressed as two affine coordinates $x$ and $y$ on an EC. The result of the multiplication is another point belonging to the same EC as $P$. The security of EC protocol is guaranteed only as long as the private key is kept secret. If a device processes an authentication request, it performs a $kP$ operation, where the scalar $k$ is its private key. If a device generates a signature, it performs a $kP$ operation using a random number $r$ as the scalar $k$. Nevertheless, revealing of scalar $r$ allows the adversary to easily calculate the private key used for a signature generation. Therefore, we call scalar $k$ the *key*. The goal of an attacker is to reveal the key.

The point multiplication is a complex operation involving the coordinates of point $P$, scalar $k$ and parameters of the underlying EC. The Montgomery $kP$ algorithm using Lopez-Dahab projective coordinates [1] is a bitwise processing of scalar $k$, using field operations involving: addition, squaring and multiplication of $GF(2^n)$ elements and register bank access. Devices performing a cryptographic operation consume energy and need time for the calculation. Their current and electromagnetic emanation can be measured and analysed statistically, with the goal to distinguish the parts processing key bits '1' from the parts processing key bits '0'. Such attacks are known as side channel analysis (SCA) attacks. If the attacked device is physically accessible, which is true for WSN nodes and IoT devices, countermeasures against SCA attacks have to be implemented.

The sequence of operations in the Montgomery ladder does not depend on the processed scalar, i.e. processing of scalar bit "1" requires the same number, type and sequence of operations, as processing a bit "0". This is the reason why the Montgomery $kP$ algorithm is mentioned in the state-of-the-art literature as resistant against simple SCA attacks. However, it was shown in [2] that the register operations, in particular the register addressing, can cause SCA leakage which in turn can be successfully used for revealing the secret scalar $k$. In [3] a horizontal address bit Differential Power Analysis (DPA) was performed using a single power trace of a $kP$ execution. The assumption behind the attack is that, although the sequence and type of operations are the same, regardless of the processed bit of scalar *k*, the power consumption of the addressing of registers depends on the key bit value and is distinguishable.

This work is part of a larger research initiative focusing on two goals. The first one consists in developing methods to uncover vulnerabilities in *kP* designs, i.e. finding SCA leakage sources. The second one – in implementing countermeasures and creating SCA resistant *kP* designs. If SCA leakage sources are detected at early stages of the designing process, this knowledge can be used for re-designing the implementation. Thus, the effective SCA methods can be a part of the efficient design methodology.

In this work we compare two different methods for the analysis of power or electromagnetic traces of *kP* executions. The first method is *the comparison to the mean* [4], which is a simple statistical analysis method. The second one is one of the mostly used unsupervised machine learning algorithms for data clustering – *K-means* [5]. The results obtained in our early work [6] show that *the comparison to the mean* was even better than the machine learning algorithm at revealing the secret key. In this work we concentrate on the comparison of the attack results using both analysis approaches with the goal to understand the benefits of each of them.

The structure of the rest of this paper is as follows. In section II we describe how we obtained power traces of *kP* executions for our experiments. In section III the preparation of the traces as well as the methods applied for their analysis are explained. Attack results are given and compared in section IV. The paper ends with short conclusion.

## II. DATA ACQUISITION

The design investigated here is the same as Design1 in [6]. We do not describe here implementation details due to the fact, that we are not discussing implementation aspects, but concentrate on comparing the methods for the analysis of traces of *kP* executions. In order to ensure a fair comparison we analysed the same simulated power traces using two different analysis methods – *the comparison to the mean* and *K-means*.

The *kP* design used in this work is a modification of the Montgomery *kP* algorithm. The scalar $k = k_{n-1}k_{n-2}k_{n-3} \ldots k_1k_0$ is processed from its most significant bit $k_{n-1}$ to its least significant bit $k_0$. The first two bits: $k_{n-1}$ and $k_{n-2}$ are processed before the main loop starts. This prevents the revealing of bit $k_{n-2}$ by simple analysis and increases the resistance against template attacks if the randomization of projective coordinates of the EC point *P* is not implemented. The length *n* of *k* is limited by the order of the EC, which in this case is NIST Elliptic Curve *B-233* [7], posing a constraint on all factors of *kP* operation to be maximally $n_{max}$=233 bits.

We synthesized the design for the IHP 250nm technology [8] using Synopsys Design Compiler Version K-2015.06-SP2. The power simulation was done using Synopsys' PrimePower Version Q-2019.12-SP1 for linux64. The clock cycle was set to 30 nanoseconds and the waveform time step for simulation to 0.1 nanoseconds resulting in *S=300* samples per clock cycle.

We concentrated only on the processing of key bits in the main loop of the implemented *kP* algorithm that corresponds to the processing of (*n-2*) bits of the scalar *k*. In [6] the hamming weight (number of ones) of this part ($k_{n-3} \ldots k_1k_0$) was *d = 121*, and its length $l = n-2 = 230$, resulting in 53% of '1'-values. We call keys balanced if they consist of almost the same number of ones and zeroes.

In [6] we applied *the comparison to the mean* and *K-means* for the analysis of *kP* traces representing the processing of a balanced key. In this work we evaluated the performance of these methods in relation with different hamming weights of the key. We assume that *K-means* is the better alternative if the scalar *k* is unbalanced. Our goal is to evaluate this assumption.

We have simulated 27 *kP* executions, always using the same base point *G* for the NIST EC *B-233* (for the coordinates of the point see [7]). The hamming weight of the scalar processed was increased from 0 to 230 in variable steps. Smaller steps were taken when the results changed significantly between executions. Table I. shows the hamming weights of the keys for each of the *kP* executions, *d* denotes the hamming weight of the key and *d/n [%]* is the percentage of ones.

TABLE I.  HAMMING WEIGHT *d* OF SCALARS PROCESSED IN ANALYSED *kP* EXECUTIONS

| No. | *d* (*d/n*, %) | No. | *d* (*d/n*, %) | No. | *d* (*d/n*, %) |
|---|---|---|---|---|---|
| 1 | 0 (0) | 10 | 41 (18) | 19 | 193 (84) |
| 2 | 4 (2) | 11 | 46 (20) | 20 | 197 (86) |
| 3 | 9 (4) | 12 | 69 (30) | 21 | 202 (88) |
| 4 | 13 (6) | 13 | 92 (40) | 22 | 207 (90) |
| 5 | 18 (8) | 14 | 115 (50) | 23 | 211 (92) |
| 6 | 23 (10) | 15 | 138 (60) | 24 | 216 (94) |
| 7 | 27 (12) | 16 | 161 (70) | 25 | 220 (96) |
| 8 | 32 (14) | 17 | 184 (80) | 26 | 225 (98) |
| 9 | 36 (16) | 18 | 188 (82) | 27 | 230 (100) |

Each execution results in one simulated power trace. Each power trace is a record of processing *l=230* bits in the main loop. Each iteration of the loop is *m=54* clock cycles long. These *m* clock cycles associated with a single iteration are called a slot, i.e. each slot corresponds to the processing of a key bit value. Each analysed power trace contains $l \cdot m \cdot S = 230 \cdot 54 \cdot 300 = 3726000$ samples.

### III. METHODS APPLIED FOR POWER ANALYSIS

In this work we concentrate on two methods that we applied in our horizontal attacks:

- *comparison to the mean*
- *K-means* clustering

A short description of the attack methods and preparation of traces for the attacks is given below.

#### A. Comparison to the mean

*The comparison to the mean* method [4] is based on the assumption that the profiles of the processing of different key bits can be distinguished by statistical methods applied to a single trace. The trace is represented as a sequence of *j* parts. Each part, further denoted as a slot, corresponds to the processing of a single key bit. A slot consists of *i* compressed power values drawn from the power supply during the processing of the current bit. The raw trace is compressed to simplify the analysis of the traces. We represent each clock cycle using a single value. There are many different compression techniques. In this paper, we calculated a sum of power values within a clock cycle and used this sum to represent the clock cycle. It can be summarized as:

$$x_{j,i} = \sum_{s=1}^{S=300} v_{j,i,s} \quad (1)$$

where *j* denotes slot number in the trace ($n-2 \geq j \geq 0$), *i* denotes a clock cycle number in the $j^{th}$ slot ($1 \leq i \leq 54$) and *s* – a sample number in the $i^{th}$ clock cycle ($1 \leq s \leq 300$).

For each *i* we calculate an average power consumption within the whole trace as follows:

$$\overline{x}_i = \frac{\sum_{j=0}^{l=n-2} x_{j,i}}{l} \quad (2)$$

For each of the *j* slots the value that represents its $i^{th}$ clock cycle, i.e. the value $x_{j,i}$ is compared against the corresponding average value $\overline{x}_i$. If $x_{j,i}$ is higher than or equal to the average, we assumed that this slot corresponds to the processing of a key bit value 'one', otherwise – of the key bit value 'zero'. The classification process can be described as:

$$\hat{k}_{i,j} = 1 \text{ if } x_{j,i} \geq \overline{x}_i, \quad \text{otherwise } \hat{k}_{i,j} = 0 \quad (3)$$

This method results in *i* key candidates $\hat{k}_i$, i.e. one per clock cycle.

In our implementation the processing of a key bit takes 54 clock cycles, whereby each clock cycle is represented using only a single compressed power value. Therefore, each slot consists of 54 samples and the attack results in 54 key candidates.

#### B. K-means

*K-means* is a well-known iterative clustering algorithm. Its application in SCA attacks is described in detail in [6]. We describe *K-means* here without using formulae with the goal to improve the readability of this paper. We used the *K-means* method to assign the *l = (n - 2)* slots of each simulated power trace into one of two clusters: one representing the processing of key bits '1' and the second one representing the processing of key bits '0'. In our approach we represent each trace containing *l*=230 slots (each slot contains 54 samples) as 54 vectors of length *l*=230. The current number of the vector is denoted with *i* ($1 \leq i \leq 54$).

Each vector is analysed separately, therefore for each vector the algorithm is initialized with K centroids. A centroid at the initialization phase is simply a random point in the vector. After the algorithm finished there will be K clusters of points. Since we want to distinguish slots corresponding to either zero or one, we set K=2. The algorithm computes the distance between each point in the vector and both centroids. The distance is the Euclidian distance, also known as L2-norm. Points laying closer to the first centroid are assigned to the first cluster and points laying closer to the second centroid are assigned to the second cluster.

Afterwards an average of all points within a cluster is calculated. This average is the new centroid of the cluster (an artificial point). The assignment to clusters is deleted and all points are clustered again based on their distance to the new centroids. The algorithm repeats the clustering until the centroids no longer move or a maximum number of iterations is met. In our case 300 iterations are allowed. Since the initialization is random, the algorithm is initialized 10 times and the result with the smallest inertia is chosen. Inertia is the sum of squared distances between all points in the cluster to its centroid. It says how "tight", or "condensed" the cluster is.

When the processing according to the algorithm is finished all points in the vector belong to one of two clusters. The centroids of those clusters mark the average of all points in the cluster. We assume that the first cluster represents slots processing a one, and the second cluster – zeros. Because there are 54 vectors, there are 54 key candidates, i.e. we obtained the same number of the key candidates as using *the comparison to the mean* method.

### IV. EVALUATION OF ATTACK RESULTS

To evaluate the success of the attack, we compared the obtained 54 key candidates with the scalar processed during the attacked *kP* execution for each of the two analysis methods. We also take into account that the processing of a single key bit goes beyond a single iteration of the main loop, i.e. some operations in slot number *i* belong to the processing of key bit *i+1* or *i-1*. Therefore we compare each revealed key candidate with the scalar shifted one bit to the left and one bit to the right. The correctness of each key candidate was calculated as the number of correctly revealed key bits, divided by the number of key bits processed in the main loop of the attacked algorithm, i.e. 230.

Figure 1 shows the correctness of 54 key candidates when a balanced key was processed. The results for both methods were quite similar.

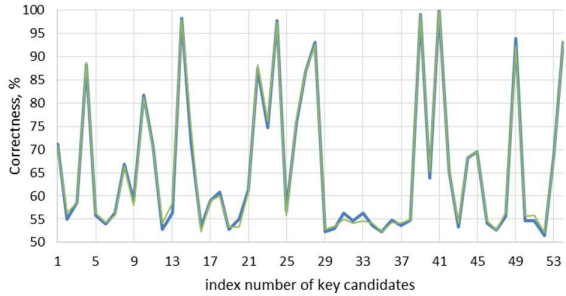

Figure 1.  The correctness of the keys revealed when analysing a power trace simulated for a balanced key: *the comparison to the mean* attack (blue line) and *K-means* (green line).

Both methods are able to reveal the processed scalar completely: the correctness of the 41st key candidate is 1 (or 100%). This means that the 41st key candidate is equal to the processed key $k$. Additionally, there are 3 key candidates – 14th, 24th, 39th – extracted with a correctness of more than 95%. The correctness of the key candidates 28, 49 and 53 is about 93%.

*The comparison to the mean* works properly if the scalar $k$ is balanced. Due to the nature of *the comparison to the mean* method, where the *mean slot* will be compared with each slot. The *comparisons to the mean* cannot work properly if the processed scalar $k$ contains mostly 'ones' or mostly 'zeros'. Thus, the applied methods can provide significantly different results for analysing traces processing unbalanced keys, i.e. if the number of key bits 'one' is significantly higher or smaller than the number of key bits 'zero'. Thus, we expected that the *K-means* method is by far more effective in case of an unbalanced key. In order to evaluate this assumption, we performed horizontal attacks using both analysis methods against 27 simulated power traces.

Figure 2 shows the overall results of the correctness of the keys revealed for both methods. For each of the 27 investigated cases only the key candidate with the maximum correctness was chosen. The Hamming weight of the key grows from left to right, with a balanced key in the middle (marked with a dashed line).

Both methods perform similarly well when the key contains between 40% and 60% of ones. When the number of ones is smaller than 30%, both methods fail to reveal even one candidate with correctness of at least 95%, whereas *K-means* achieves significantly better correctness than *the comparison to the mean*. For the keys with 70% to 98% of 'ones', *the comparison to the mean* stops giving reliable results while *K means* continue to provide some key candidates with a correctness of about 100%. Both methods failed to reveal the key consisting of only ones, which is not a surprising result.

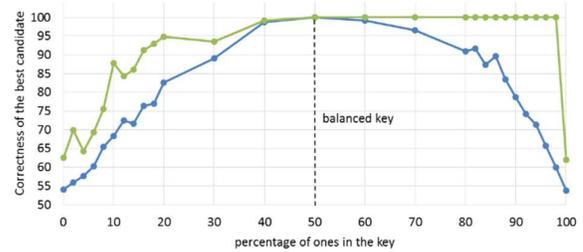

Figure 2.  The relation between the correctness of the best key and the Hamming weight of the key. The results are obtained by attacking the simulated power traces using *the comparison to the mean* (blue line) and *K-means* (green line).

In order to ensure a fair assessment of our results, we checked the correctness not only for the best key candidate out of the 54 but also for 3 additional key candidates. The selected key candidates are candidate 22, 41 and 49, i.e. those with the indexes $i$ = 22, 41 and 49. Their correctness is close to 90%, 100% and 95% respectively in case of a balanced key (see Figure 1). Figure 3 and 4 show the correctness of these three key candidates for *the comparison to the mean* and *K-means* for all the different hamming weights used. To simplify comparison the best correctness for each method shown in Figure 2 is given also in Figure 3 and 4, see line denoted "best out of 54".

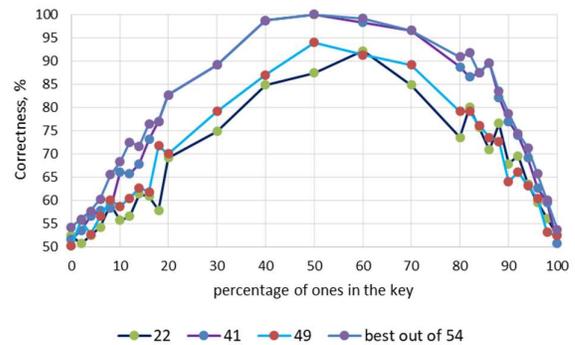

Figure 3.  The correctness of the key candidates 22, 41 and 49 vs. hamming weight of the processed keys for *the comparison to the mean* method; the line "best out of 54" and the blue line in Figure 2 are the same.

*K-means* provides more stable results for unbalanced keys than *the comparison to the mean* approach, i.e. *K-means* is more effective when attacking traces representing the processing of unbalanced scalars. *K-means* performs better because the clustering method does not depend on the number of elements that belong to each cluster. It is still effective even if one of the clusters contains only a few elements.

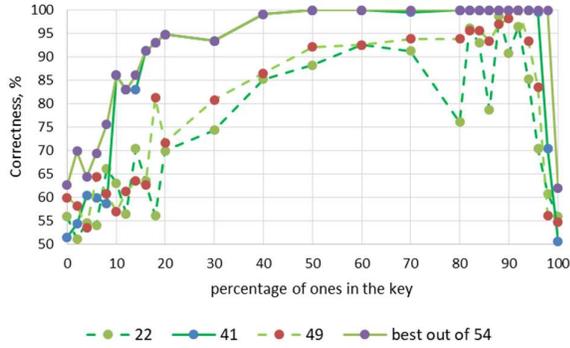

Figure 4. The correctness of the key candidates 22, 41 and 49 vs. hamming weight of the processed keys for the *K-means* method; the line "best out of 54" and the green line in Figure 2 are the same.

V. CONCLUSIONS

In this work we compared the success rate of two methods – *the comparison to the mean* and *K-means* – that can be used to reveal a scalar *k* by analysing power traces. In this paper we used simulated power traces of *kP* executions. The success of the attack depends on the applied analysis method and on the key processed during the attacked cryptographic operation. *The comparison to the mean* is a simple and very fast method for horizontal SCA attacks. It works properly if the scalar *k* processed consists of 40-60% of 'ones' bits. *K-means* is beneficially if the scalar *k* is dominated by 'ones' or by 'zeros', even if the number of 'ones' is higher than 90%. Thus, the success of SCA attacks performed using *K-means* method does not depend on the key processed during the attacked cryptographic operation. Due to this fact, *K-means* is an attractive method for evaluating the resistance of a design against SCA during its implementation.